\documentclass{ctr_summer}

\usepackage{ctrfont}
\usepackage{natbib}
\usepackage{undertilde}
\usepackage{color}

\usepackage{graphicx}
\usepackage{psfrag}
\usepackage{epsfig}
\usepackage{amsmath,rotating}
\usepackage{dashrule}
\usepackage{undertilde}

\usepackage{url}

\usepackage{color}

\usepackage{makecell}

\usepackage[titletoc,title]{appendix}

\newcommand{\vecq}{\mathbf{q}}

\providecommand\udelta{\delta}

\newcommand\lf{\left }
\newcommand\rg{\right }

\title{Symmetry breaking in a 3D bluff-body wake}
\shorttitle{Symmetry breaking in a 3D bluff-body wake}

\author{G. Rigas\footnote[1]{California Institute of Technology}, L. Esclapez \and L. Magri}

\shortauthor{Rigas et al.}

\begin{document}

\setcounter{page}{1}

\maketitle

The dynamics of a three-dimensional axisymmetric bluff-body wake are examined at low Reynolds regimes where transitions take place through spatio-temporal symmetry breaking. A linear stability analysis is employed to identify the critical Reynolds number associated with symmetry breaking, and the associated eigenmodes, known as global modes. The analysis shows that the axisymmetric stable base flow breaks the rotational symmetry through a pitchfork $m=1$ bifurcation, in agreement with previously reported results for axisymmetric wakes. Above this threshold, the stable base flow is steady and three-dimensional with planar symmetry.  A three-dimensional global stability analysis  around the steady reflectionally symmetric base flow, assuming no homogeneous directions, predicts accurately the Hopf bifurcation threshold, which leads to asymmetric vortex shedding. DNS simulations validate the stability results and characterize the flow topology during the early chaotic regime. \\

\hrule

\section{Introduction}

Bluff-body flows are of fundamental importance to many industries, in particular the transport industry, where the aerodynamic drag arising from such flows can be the dominant source of vehicle fuel-burn and $\text{CO}_2$ emissions. Recent advances in hydrodynamic stability have further aided understanding and controling laminar and transitional regimes \citep{sipp2010dynamics,Luchini2014,sipp2016linear}, particularly  for two-dimensional (2D) flows. However, flows of practical and industrial interest involve three-dimensional (3D) wakes and high Reynolds numbers. Despite their turbulence and complexity, such flows exhibit organization, which manifests as coherent flow structures. These structures are usually associated with increased noise, structural fatigue and drag. 

During the transitional regime of laminar wakes, continuous spatial and temporal symmetries are spontaneously broken through a sequence of bifurcations. Specifically, the axisymmetric 3D wake undergoes a steady bifurcation followed by an unsteady one  with azimuthal wavenumbers $|m|=1$ at low Reynolds numbers prior to the emergence of chaos \citep{fabre2008bifurcations,meliga2009global,Bohorquez2011,bury2012transitions}. These bifurcations break spatial-rotational and time-translation symmetries, giving rise to a reflectionally symmetric steady flow and unsteady vortex shedding.  Similar transitional behavior has also been observed in 3D bluff bodies with reflectional symmetries, such as square plates \citep{Marquet2015}.

Perhaps surprisingly, it was shown recently that the turbulent dominant dynamics of 3D wakes can be linked to the hydrodynamic instabilities observed during the transitional regimes at low Reynolds numbers, including axisymmetric geometries \citep{Rigas2014JFMr,Rigas2015JFMr} and square ones, such as the Ahmed body \citep{grandemange2013turbulent}. Understanding the wake dynamics in the laminar and transitional regimes is paramount for the development of control strategies, which can be appropriately extended to turbulent cases \citep{BrackstonJFM2016}.

In this study, global linear stability analysis (LSA) and direct numerical simulation (DNS) are employed to study the intrinsic dynamics leading to chaotic/turbulent behavior  in a 3D axisymmetric bluff-body wake. The remainder of this paper is structured as follows. Section~2 briefly outlines the computational set-up. Section~3 presents the predictions of the LSA. Section~4 describes the transitional regimes obtained from DNS. Finally, conclusions are drawn in Section~5. 

\section{Computational setup}

The geometry employed in this study is a 3D axisymmetric bluff body (Figure~\ref{fig:DNS_geom}) with a blunt trailing edge, which has been studied experimentally in \cite{Rigas2014JFMr,Rigas2015JFMr} and \cite{oxlade2015high}. The length-to-diameter ratio, $L/D$, is 6.48. The bluff-body nose employs a modified super-ellipsoid with an aspect ratio $AR=2.5$, given by the revolution of $y=(1-z/AR)^{m} + y^2=1$, $0<z<AR$, $m=2+(z/AR)^2$ around the centerline \citep{Lin1992}. Cylindrical ($r$, $\theta$, $z$) and Cartesian ($x$, $y$, $z$) coordinates with the origin taken at the center of the base of the body are used in the subsequent analysis. 

\subsection{Stability calculations}
The fluid motion is governed by the incompressible Navier-Stokes (NS) equations
\begin{equation}
\partial_t \mathbf{u} + \mathbf{u} \cdot \nabla \mathbf{u} +\nabla p - Re^{-1} \nabla^2 \mathbf{u}  =0 	
\mathrm{,}~~~~~~
 \nabla \cdot  \mathbf{u}=0.
\label{eq:NS}
\end{equation}
In the stability analysis, $\mathbf{u}=(u_r,u_\theta,u_z)^T $ where $u_r$, $u_\theta$, $u_z$ are the radial, azimuthal and axial components of the velocity. The stability analysis examines the evolution of infinitesimal perturbations around fixed-point solutions of \eqref{eq:NS}, known as base flows.

The base flow $\mathbf{q}_0=(\mathbf{u}_0, p_0)^T$ is a steady solution of the NS equations
\begin{equation}
\mathbf{u_0} \cdot \nabla \mathbf{u_0} + \nabla p_0 - Re^{-1} \nabla^2 \mathbf{u_0}  =0
~~\mathrm{and}~~
 \nabla \cdot  \mathbf{u_0}=0.
 \label{eq:base flow}
\end{equation}

The linear perturbation equations read
\begin{equation}\label{eq:lin}
\partial_t \mathbf{u}' + \mathbf{u_0} \cdot \nabla \mathbf{u}' + \mathbf{u}' \cdot \nabla \mathbf{u_0}+\nabla p' - Re^{-1} \nabla^2 \mathbf{u}'  =0
~~\mathrm{and}~~
 \nabla \cdot  \mathbf{u}'=0,
\end{equation}
which, in compact form, read
\begin{equation}
\mathcal{B}\partial_t \mathbf{q}' + \mathcal{A}\mathbf{q}'   =0.
\end{equation}
Assuming no homogeneous directions, the solutions are sought as normal modes with linear growth rate $\sigma$ and frequency $\omega$
\begin{equation}
\mathbf{q}' = \hat{\mathbf{q}}_{3D} (r,\theta,z) e^{(\sigma +i\omega)t} + c.c.
\end{equation}
 For an axisymmetric base flow, the azimuthal coordinate can be further Fourier-transformed in terms of the azimuthal wavenumber $m$, where $m$ is an integer parameter, as
\begin{equation}
\mathbf{q}' = \hat{\mathbf{q}}_{2D} (r,z) e^{(\sigma +i\omega)t + im\theta} + c.c.
\end{equation}
Substituting the normal-mode decompositions into Eq.~\eqref{eq:lin} leads to 3D and 2D generalized eigenvalue problems with appropriate boundary conditions. These read
\begin{equation}
(\sigma +i\omega)\mathcal{B} \hat{\mathbf{q}} + \mathcal{A}_{(m)}\hat{\mathbf{q}}   =0.
\end{equation}

For each global mode $\hat{\mathbf{q}}$ the correspondent adjoint mode $\hat{\mathbf{q}}^\dagger$ is computed from 
\begin{equation}
(\sigma -i\omega)\mathcal{B}^\dagger \hat{\mathbf{q}}^\dagger + \mathcal{A}_{(m)} ^\dagger \hat{\mathbf{q}} ^\dagger  =0. 
\end{equation}
 The adjoint operators can be either derived analytically from the direct equations (Eq.~\eqref{eq:lin}),
which is known as the continuous adjoint,
or formed from the numerically discretized operator,
which is known as the discrete adjoint \citep{Luchini2014}.
For this flow configuration and the numerical schemes adopted, the discrete-adjoint formulation is adopted. 
Hence, the adjoint operators are given by the complex conjugate of the discretized direct operators. 
The adjoint equations can be used to evaluate the effects of generic initial conditions and forcing terms on the time-asymptotic behavior of the system. Moreover, once the direct and adjoint modes are computed, the sensitivity of the global modes to local feedback (structural sensitivity,  \cite{giannetti2007structural}) and base-flow modifications \citep{Bottaro2003,marquet2008sensitivity} can be examined. 

The spatial discretization of the base flow and the linear  equations is performed in the finite-element software FreeFem++  \citep{hecht2012}, interfaced with PETSc for parallel calculations, using a continuous Galerkin scheme with $P2-P1$ Taylor-Hood elements. Results presented here were obtained using an unstructured  mesh with approximately 500,000 tetrahedra elements.  The base-flow solution of the steady nonlinear equations \eqref{eq:base flow} is obtained using an iterative Newton method and the direct linear solver MUMPS \citep{MUMPS}. Convergence is reached when the $\mathcal{L}_2$-norm of the residual of the governing equations becomes smaller than $10^{-12}$. Eigenvalue problems (2D and 3D) are solved with SLEPc \citep{Hernandez2005} using a shift-and-invert method. 



\subsection{Direct numerical simulations}

\begin{figure}
\centering
\includegraphics[width=0.9\textwidth]{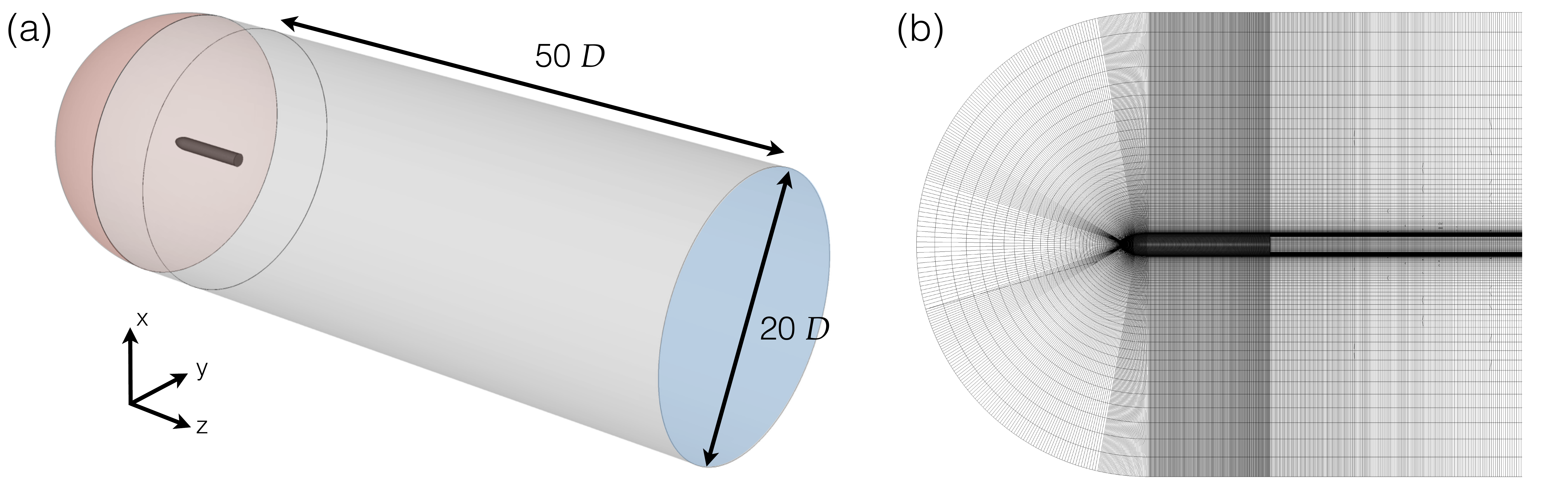}
\caption{(a) DNS computational domain. (b) Mesh in a $y$-normal plane.}
\label{fig:DNS_geom}
\end{figure}

The low-Mach-number solver {\sc Vida} (Cascade Technologies, Inc.) is used to solve the incompressible NS equations by setting constant density. It is a finite-volume, unstructured, massively parallel LES/DNS solver for reacting and non-reacting flows, fourth-order accurate in space and second-order accurate in time \citep{Ham:2007}. The equations are solved in Cartesian coordinates, with $z$ being the streamwise direction. Standard air properties at $T=20^{\circ}$ C; $\nu=15.11\times10^{-6}$ m$^2$/s; $\rho=1.205$ kg/m$^3$ are utilized. The computational domain is shown in Figure~\ref{fig:DNS_geom}(a): it consists of a 50-diameter $\times$ 20-diameter cylinder with a half-sphere inlet. A uniform velocity profile boundary condition is imposed at the inlet, no-slip at the bluff-body walls, convective at the outlet and slip conditions at the outer radial boundary. Several meshes of increasing spatial resolution were used and velocity statistics were found to converge with a mesh of about 5 million cells for $Re$ up to 1000 and of about 10 million cells for higher $Re$. The former mesh has a grid point distribution in axial, radial and azimuthal directions of $300\times50\times60$ (Figure~\ref{fig:DNS_geom}(b)), the latter has a grid point distribution of $600\times70\times120$. 
%
\section{Linear stability analysis}
\begin{figure}
	\begin{center} 
	\includegraphics[trim={0cm 13.5cm 0cm 12.5cm},clip,width=0.62\textwidth]{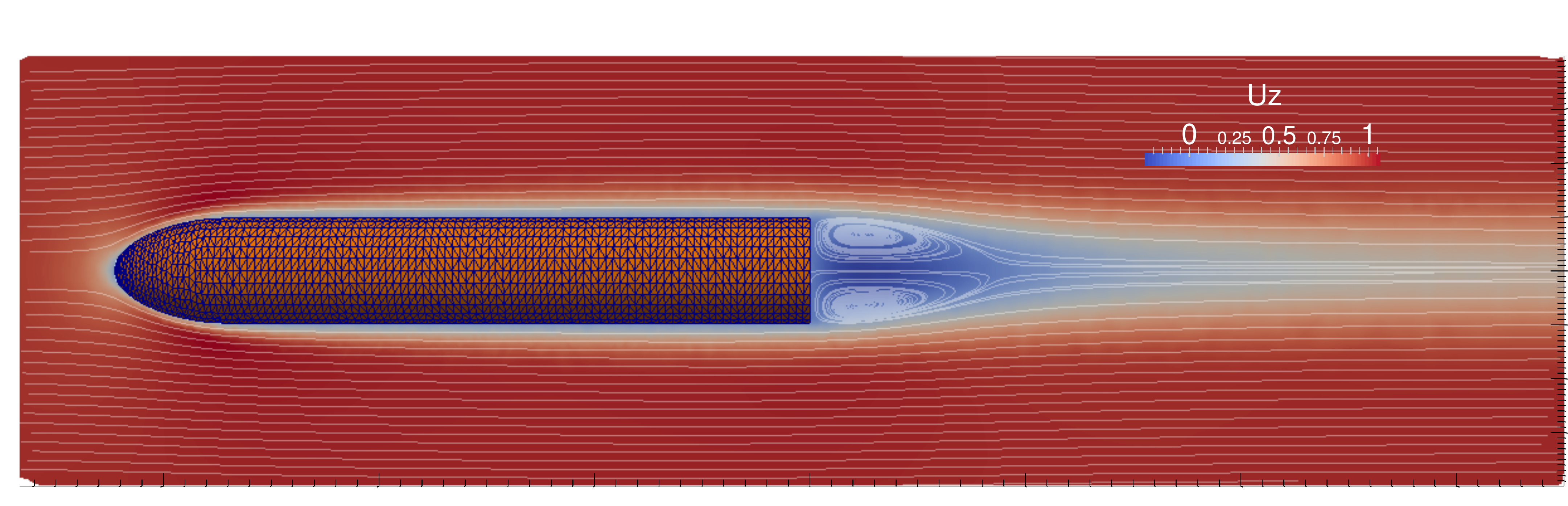}
	\includegraphics[trim={0cm 12.cm 0cm 12.5cm},clip,width=0.62\textwidth]{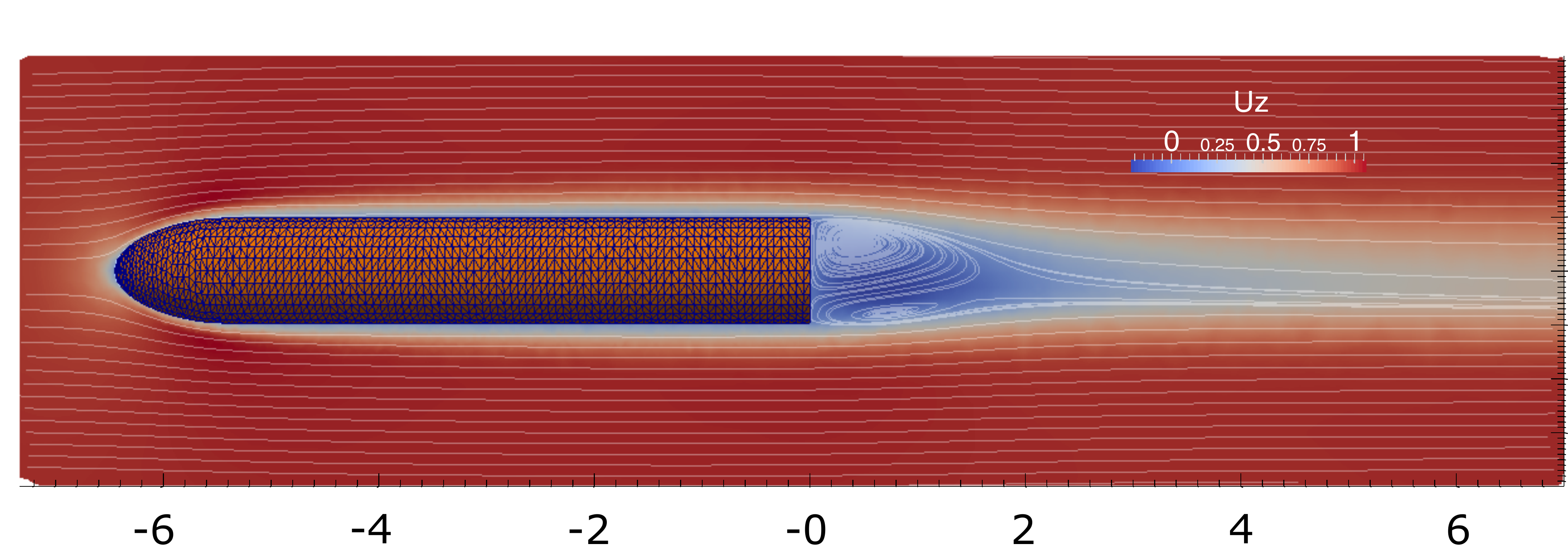}
	\caption{Base flows (fixed point solutions of the Navier-Stokes equations) at $Re=500$. Streamwise velocity component $U_z$ and streamlines in the $xz$ plane. Top: rotationally symmetric (axisymmetric)  base flow.  Bottom: reflectionally symmetric base flow with respect to the $xz$ plane.}
	\label{fig:baseflow}
	\end{center}
\end{figure}
In the limit of vanishingly small Reynolds numbers, $Re \ll 1$, the base flow is steady (time invariant) and axisymmetric (azimuthally invariant),  respecting the axisymmetry of the bluff-body configuration. The flow is stable, which means that perturbations decay asymptotically in time i.e., all the eigenvalues of the linearized NS operator have strictly negative real part, $\sigma<0$. En route to chaos, these symmetries break through a series of bifurcations that bring about reduced symmetry in space or time. The transition between the various symmetry groups is characterized here through LSA. For the flow studied in this paper, when a bifurcation occurs, the base flow (equilibrium) becomes unstable and a real eigenvalue (steady pitchfork bifurcation) or a complex conjugate pair (unsteady Hopf bifurcation) attains positive growth rates ($\sigma>0$).

\begin{figure}
	\begin{center}
	\includegraphics[width=0.49\textwidth]{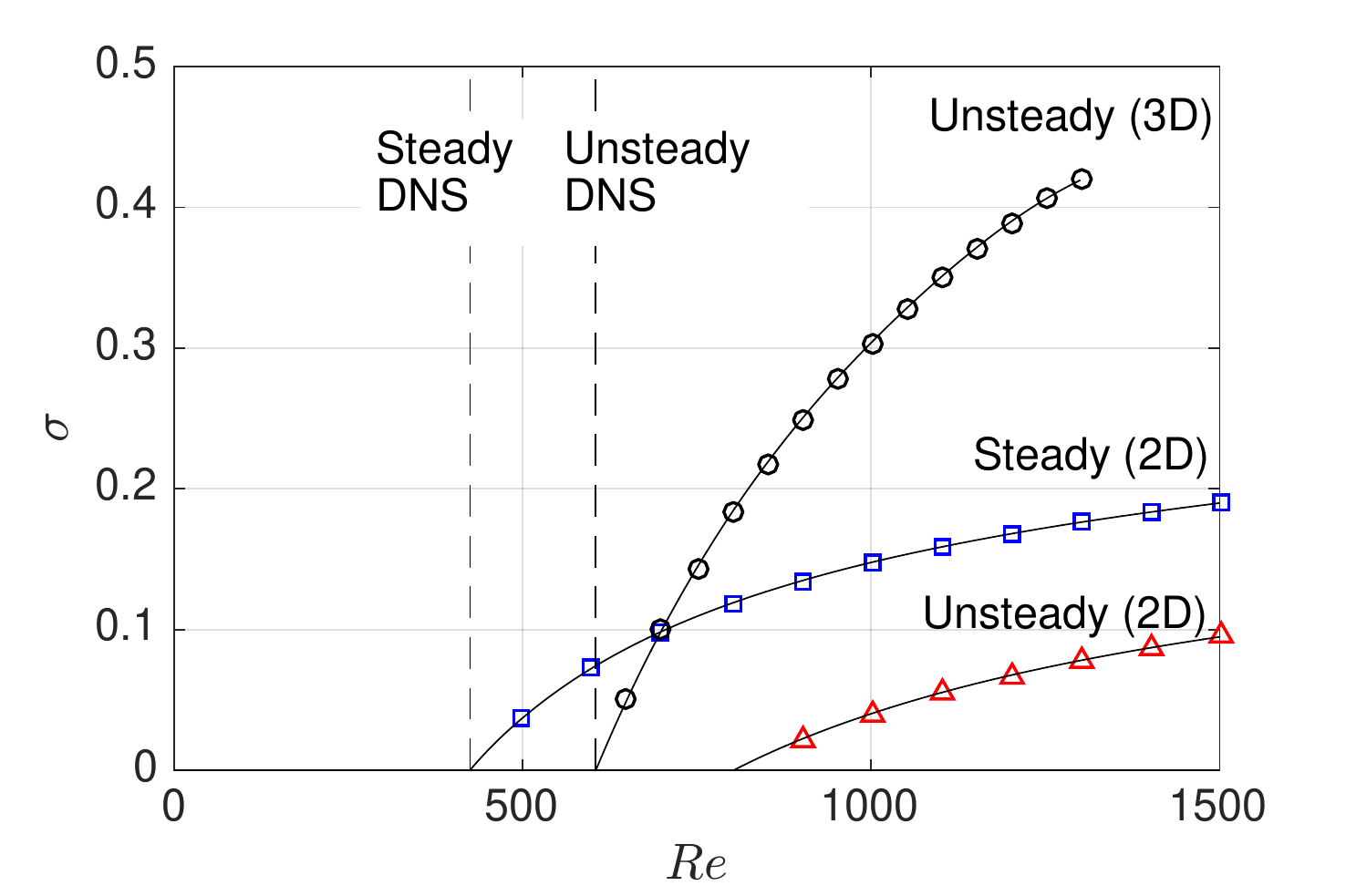}
	\includegraphics[width=0.49\textwidth]{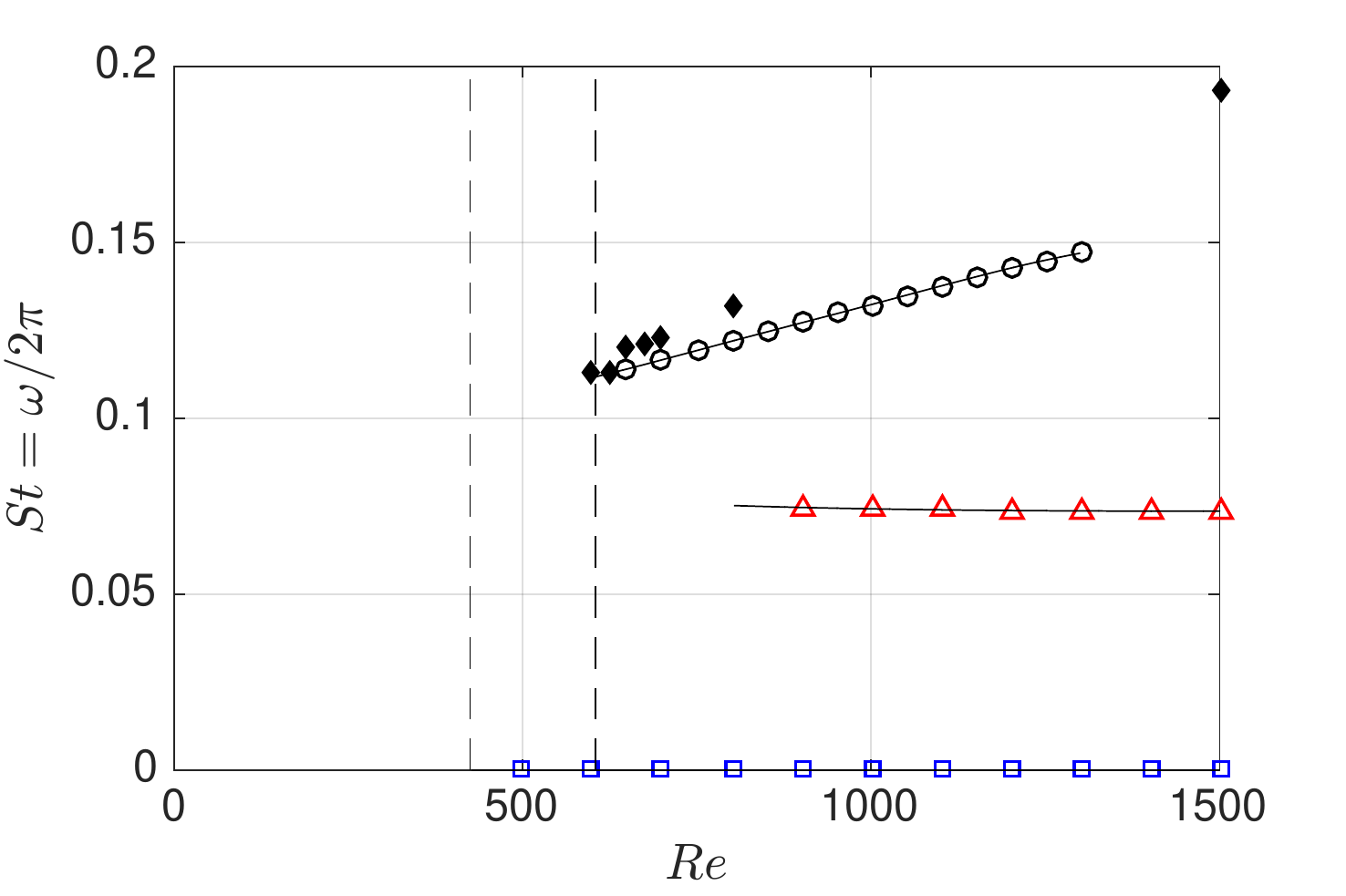}
	\caption{Growth rate (left) and frequency (right) of the most unstable global modes against Reynolds numbers. Steady mode (blue squares)  and unsteady vortex shedding mode (red triangles) predictions using the axisymmetric 2D base flow. Unsteady vortex shedding mode using the reflectionally symmetric 3D base flow (black circles). Critical $Re$ (dashed lines) and frequencies (filled symbols) from  DNS.}
	\label{fig:eig}
	\end{center}
\end{figure}

The first symmetry-breaking bifurcation is a steady bifurcation at $Re_c=424$, which breaks the rotational symmetry of the axisymmetric 2D base flow and leads to a 3D topology. Above this critical Reynolds number the axisymmetric base flow becomes unstable and a new stable base flow emerges. The stable base flow is characterized by reflectional symmetry in the azimuthal direction. These two equilibria for $Re=500$ are shown in Figure~\ref{fig:baseflow}. The growth rate and frequency obtained by examining the stability of these base flows are shown in Figure~\ref{fig:eig}.

\subsection{Global modes of axisymmetric (2D) base flow}

Two bifurcations are identified by examining the stability of the axisymmetric flow and relevant global modes.  At $Re_c=424$ the axisymmetric base flow undergoes a steady pitchfork bifurcation, with azimuthal wavenumber $m=1$ and zero frequency ($\omega_c=0$).  At $Re_c=802$ the axisymmetric base flow undergoes an unsteady Hopf bifurcation, with azimuthal wavenumbers $m=\pm1$. A complex conjugate pair of eigenvalues crosses the imaginary axis of the spectrum, with a critical Strouhal frequency of 0.075. The direct and adjoint eigenmodes for $Re=600$ are shown in Figure~\ref{fig:eigenvectors}(a,b).

The two symmetry-breaking bifurcations obtained from the stability of the axisymmetric base flow are in agreement with the findings  of \cite{sanmiguel2011effects,Bohorquez2011} for an axisymmetric body of similar geometry. The threshold for the steady bifurcation is in agreement with the critical value 420, obtained from the DNS. However, the frequency and the critical Reynolds number for the unsteady vortex shedding do not match the DNS values. This is because the axisymmetric base flow has become unstable at a lower $Re$ due to the steady bifurcation, thereby not serving as the correct base solution for predicting the Hopf bifurcation. For the same reason, this unsteady flow regime cannot be observed as a solution in the asymptotic time regime. An accurate prediction of the threshold and mode shape of the unsteady bifurcation requires the solution of the eigenvalue problem around the stable equilibrium, which is 3D, as explained in the next section.

\subsection{Global modes of reflectionally symmetric (3D) base flow}

\begin{figure}
	\begin{center} 
	\vspace{0.5cm}
	Global modes of axisymmetric base flow (2D) \\
	\includegraphics[trim={0.35cm 15.cm 0cm 15.8cm},clip,width=0.48\textwidth]{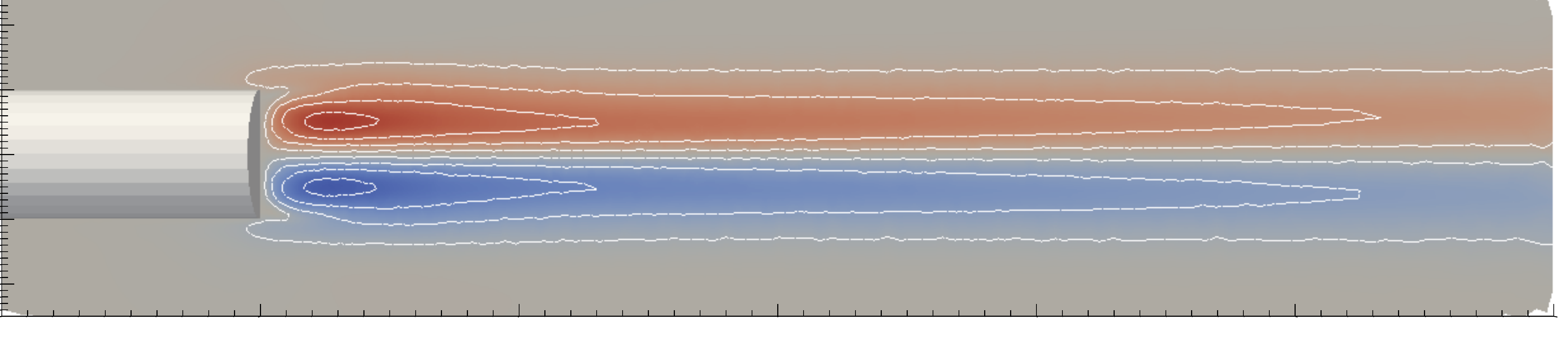}
	\includegraphics[trim={0.35cm 15cm 0cm 14.8cm},clip,width=0.48\textwidth]{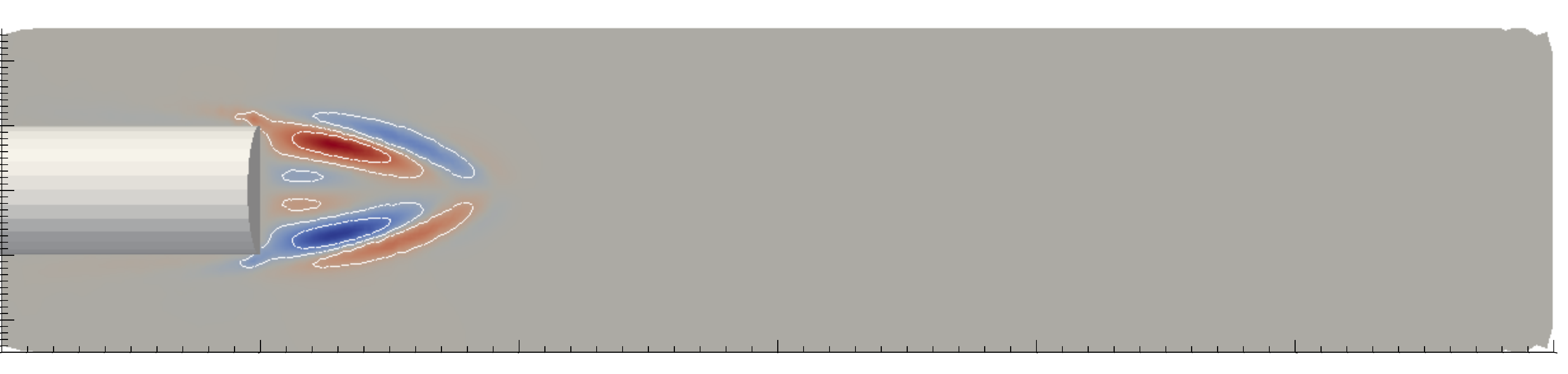}\\
	\includegraphics[trim={0.35cm 15cm 0cm 15.8cm},clip,width=0.48\textwidth]{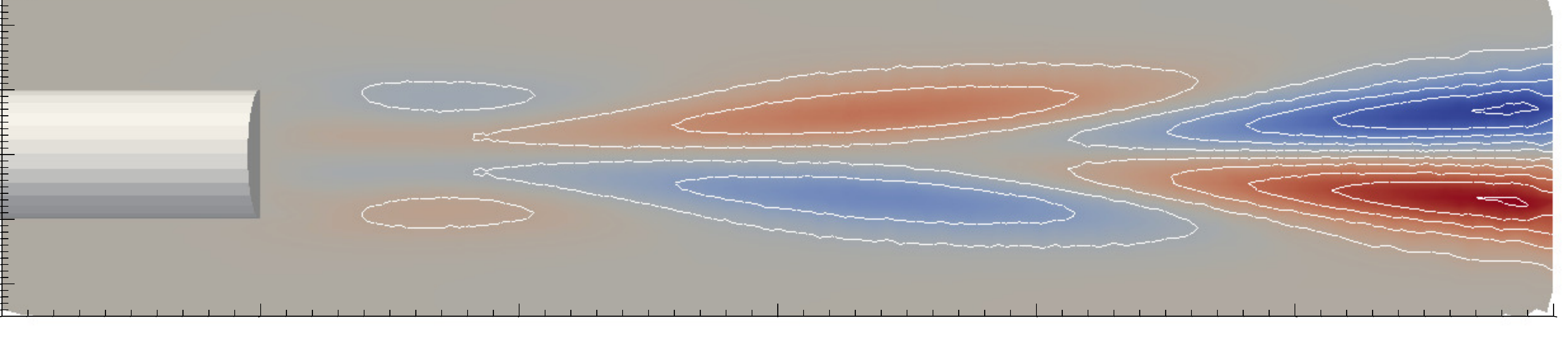}
	\includegraphics[trim={0.35cm 15cm 0cm 14.8cm},clip,width=0.48\textwidth]{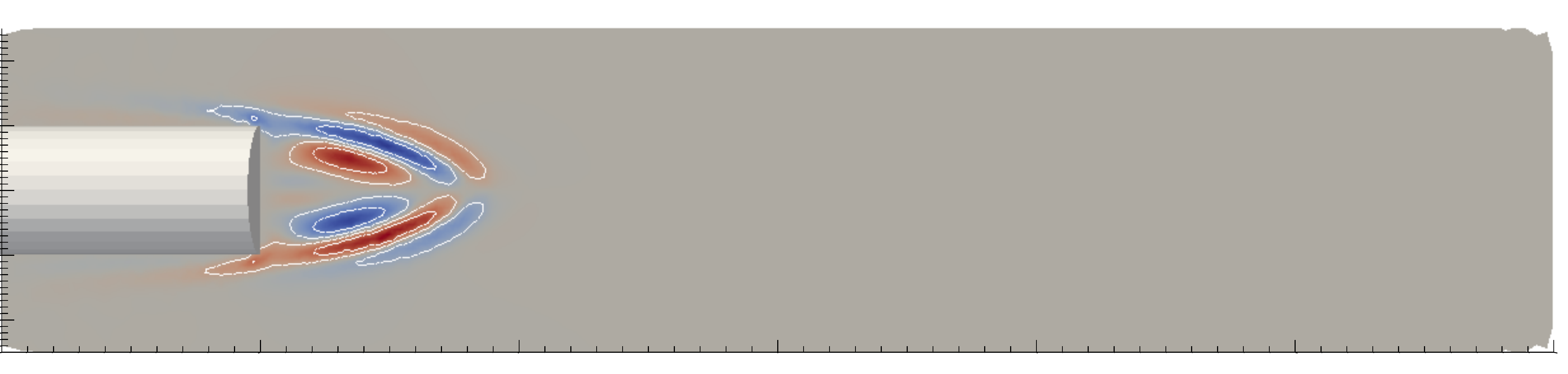}\\
	\vspace{0.2cm}
	Global modes of reflectionally symmetric base flow (3D) \\
	\includegraphics[trim={0.35cm 12.5cm 0cm 14.8cm},clip,width=0.48\textwidth]{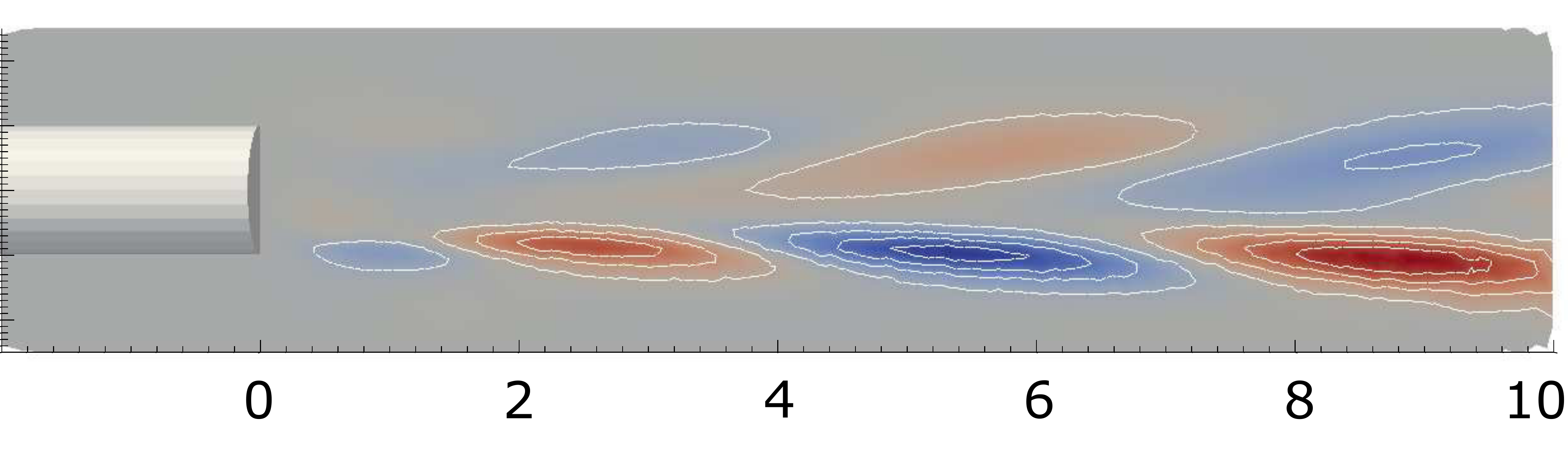}
	\includegraphics[trim={0.35cm 12.5cm 0cm 14.8cm},clip,width=0.48\textwidth]{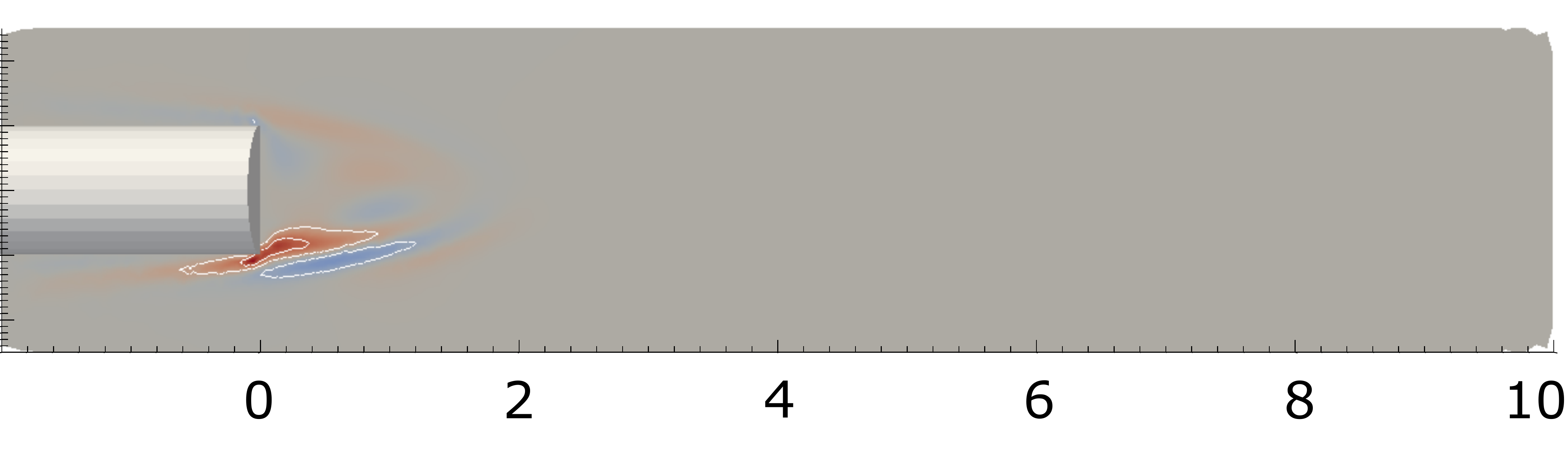}
	\caption{Global modes at $Re=600$: direct (left) and adjoint (right). Steady symmetry breaking of the axisymmetric base flow (top);  unsteady vortex shedding around the axisymmetric base flow (middle);  (c) unsteady vortex shedding around the reflectionally symmetric base flow (bottom). The real part of streamwise velocity component is shown in the $xz$ plane.}
	\label{fig:eigenvectors}
	\end{center}
\end{figure}

The new emerging stable base flow for $Re_c>424$ is characterized by reflectional symmetry and is fully 3D. We find that at $Re_c=605$ the reflectionally symmetric base flow undergoes an unsteady bifurcation. A complex conjugate pair of eigenvalues crosses the imaginary axis, with a critical Strouhal frequency of 0.113. This is in agreement with the DNS values for the threshold (600) and frequency (0.113). The associated unstable unsteady global mode inherits the symmetry properties of the base flow, as shown in Figure~\ref{fig:eigenvectors}(c), and becomes reflectionally symmetric, biased  toward the same direction as the base flow.

The LSA is valid close to the threshold of the bifurcation (right panel of Figure~\ref{fig:eig}). The frequency deviation between the DNS results and the linear stability for higher $Re$ is due to the nonlinear base-flow modification caused by the Reynolds stresses associated with the global shedding mode.  For an accurate prediction of the saturated frequency and amplitude of the global mode, a weakly nonlinear analysis  \citep{sipp2007global,meliga2009global} or a self-consistent model \citep{mantivc2014self} would be necessary. Note that the nonlinear base-flow modifications associated with the first steady mode above the threshold of the first bifurcation are fully accounted for here because of the appropriate choice of the stable reflectionally symmetric flow. For the stable and steady regimes, the axisymmetric base flow, $Re<424$, and the reflectionally symmetric base flow, $424<Re<605$, obtained by the Newton-Raphson solver, are exact solutions of the NS equations and correspond to the mean flow, as found also with DNS. The advantages of using the Newton-Raphson solver over DNS are (i) the low computational cost because the method converges after a few iterations, and (ii) the ability of obtaining unstable base flows with no further implementation. 

\subsection{Structural sensitivity}
%
\begin{figure}
	\begin{center} 
	\includegraphics[trim={3.35cm 14.0cm 25cm 14.8cm},clip,width=0.31\textwidth]{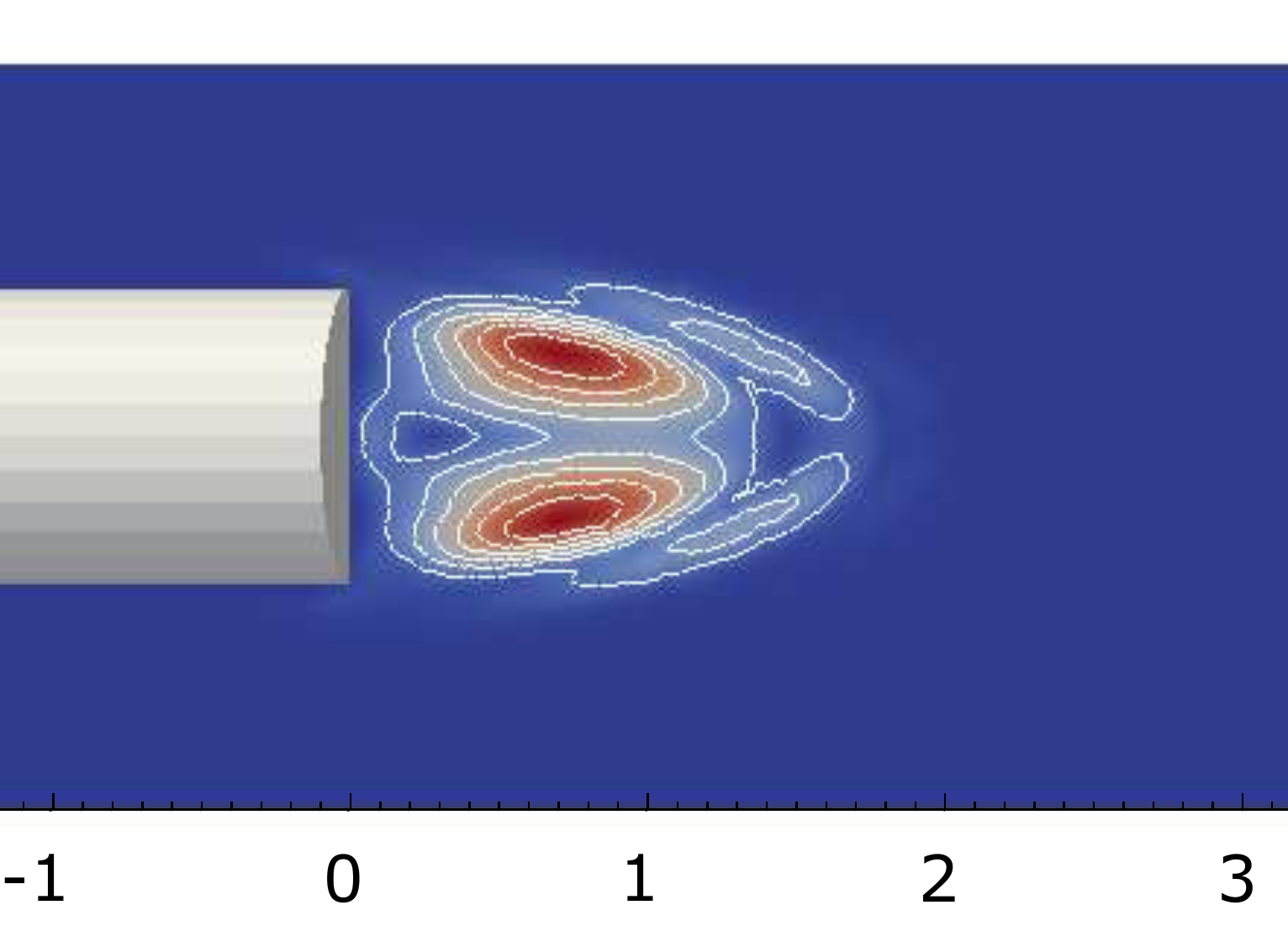}
	\hspace{0.2cm}
	\includegraphics[trim={3.35cm 14.0cm 25cm 14.8cm},clip,width=0.31\textwidth]{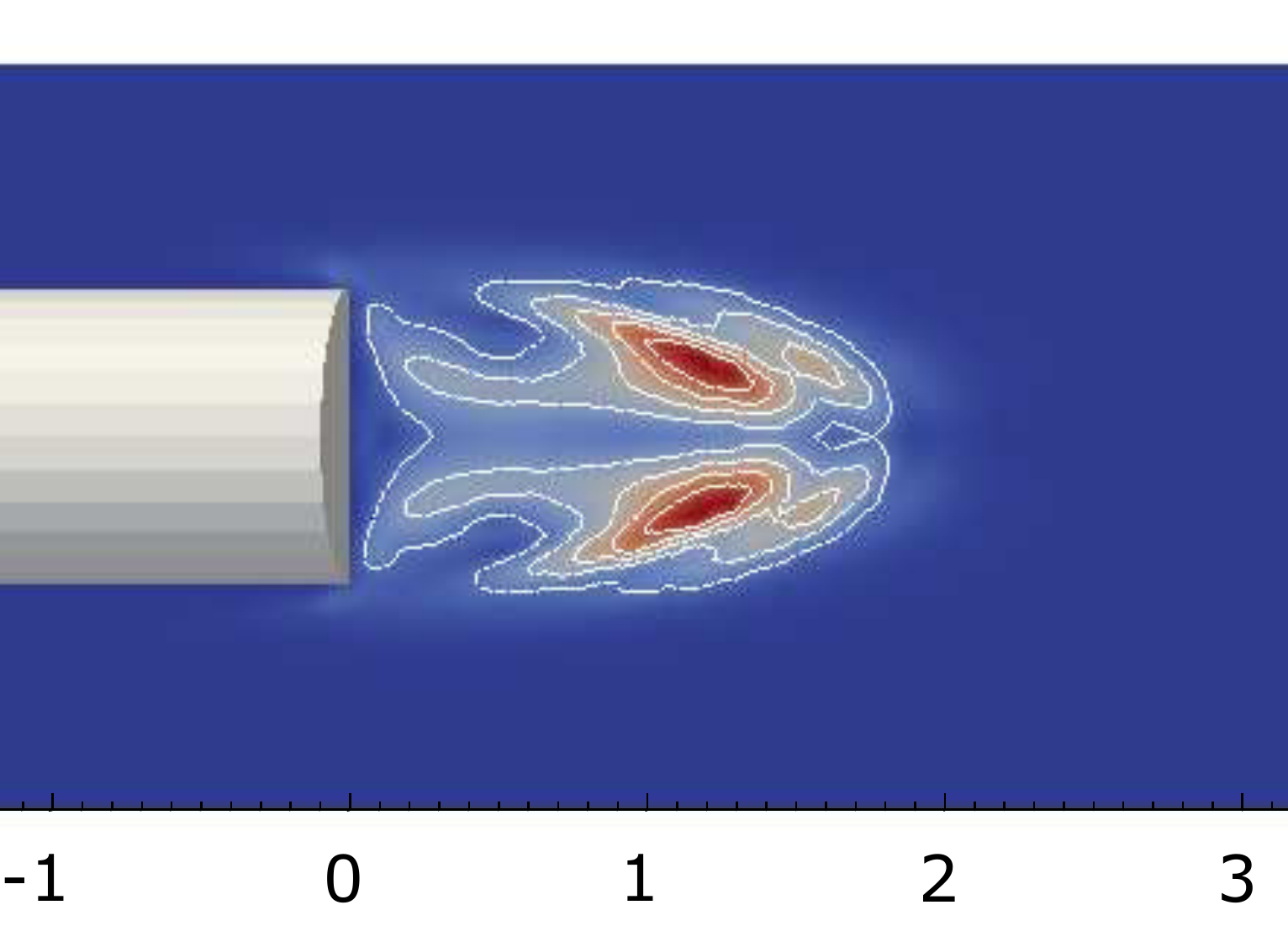}
	\hspace{0.2cm}
	\includegraphics[trim={3.35cm 14.0cm 25cm 14.8cm},clip,width=0.31\textwidth]{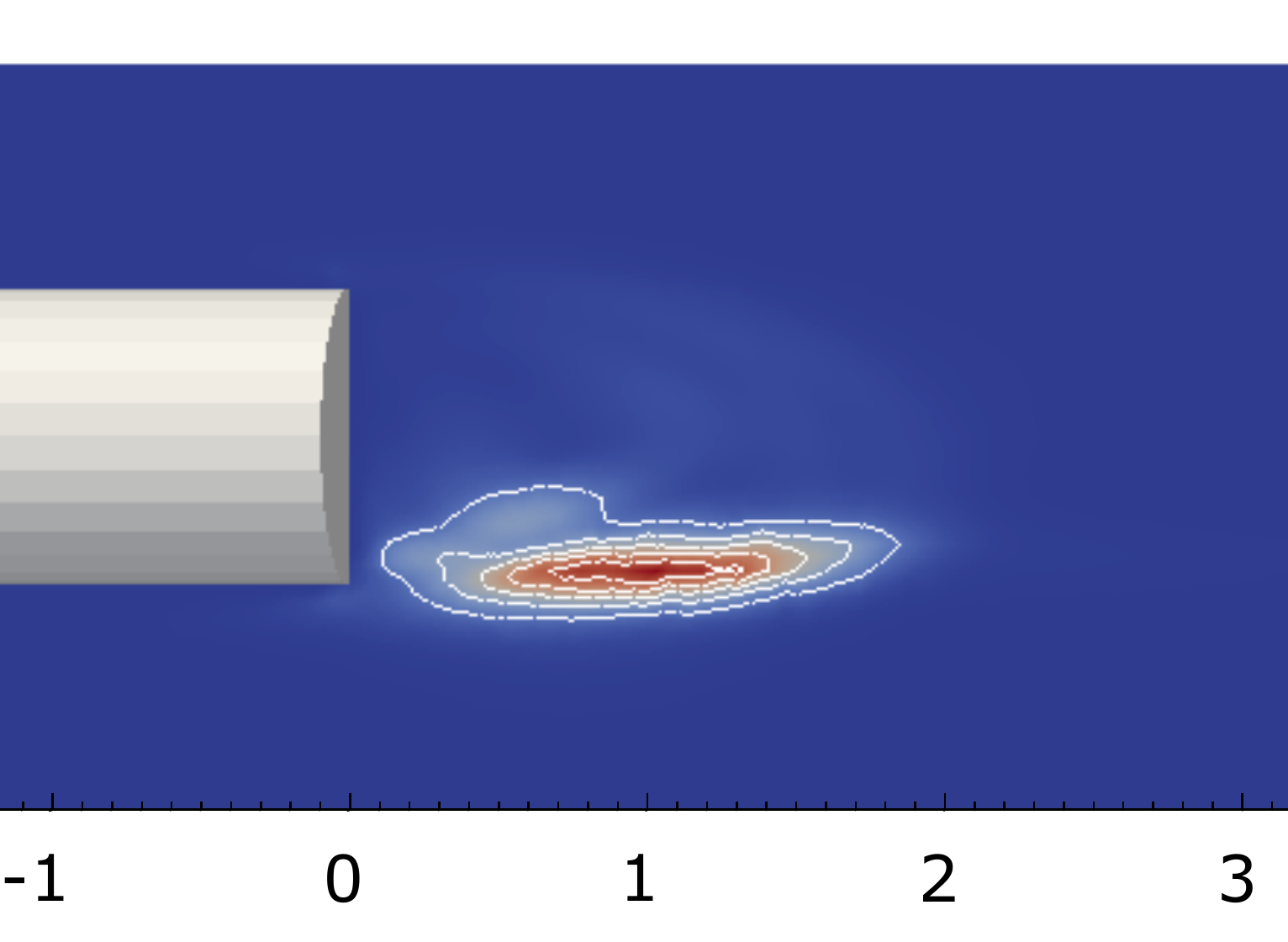}
	\caption{Structural sensitivity of the global modes shown in Figure~\ref{fig:eigenvectors}.}
	\label{fig:structural}
	\end{center}
\end{figure}
Mathematically, the structural sensitivity is related to the first-order sensitivity of an eigenvalue to small perturbations of the direct linear operator. 
When the direct operator, $\mathcal{A}$, is structurally perturbed by $\udelta\mathcal{A}$,
such that
$\mathcal{A} \rightarrow \mathcal{A} + \varepsilon  \udelta\mathcal{A} $
and
$\sigma \rightarrow \sigma + \varepsilon \udelta \sigma $, 
the first-order drift of the eigenvalue is given by the G$\mathrm{\hat{a}}$teaux derivative \citep{Magri_thesis}
\begin{equation}
 \label{eq:senstructdef}
\udelta\sigma
\equiv \lim_{\varepsilon \rightarrow 0} \lf[ \frac{ \sigma \lf( \mathcal{A} + \varepsilon\udelta\mathcal{A}
   \rg) - \sigma \lf( \mathcal{A} \rg) }{\varepsilon} \rg]. 
\end{equation}
The resulting drifts in the eigenfunctions are assumed to be perturbed as
$ \hat{\vecq}  \rightarrow \hat{\vecq} + \varepsilon \udelta \hat{\vecq}$
and
$ \hat{\vecq}^\dagger  \rightarrow \hat{\vecq}^\dagger + \varepsilon \udelta \hat{\vecq}^\dagger$.
By taking the limit $\varepsilon \rightarrow 0$
and using the bi-orthogonality condition,
it can be shown that for linear eigenvalue problems \citep[see, e.g.,][]{giannetti2007structural}
\begin{align}
\label{eq:delta_sigma_sen}
 \udelta \sigma = \frac{ \lf\langle \hat{\vecq}^\dagger, \udelta\mathcal{A} \hat{\vecq}
 \rg\rangle }{  \lf\langle \hat{\vecq}^\dagger, \mathcal{B}\hat{\vecq} \rg\rangle }, 
\end{align} 
where $\langle\cdot,\cdot\rangle$ is an inner product. (The extension of this formula for nonlinear eigenvalue problems can be found in \cite{Magri2016jcp1}). 
By considering a structural perturbation that is localized in space by the identity tensor, we define the structural sensitivity of the direct operator, $\mathcal{A}$, as \citep{giannetti2007structural}
$$
S(x,y,z) = \frac{\| \hat{\mathbf{u}}(x,y,z) \| \cdot \| \hat{\mathbf{u}} ^{\dagger}(x,y,z) \|}{  \lf\langle \hat{\vecq}^{\dagger}, \mathcal{B}\hat{\vecq} \rg\rangle }.
$$
Physically, the region in the flow acting as a wavemaker in the excitation of the global instability can be identified by considering the structural sensitivity of the unstable mode. This concept, although typically used for oscillatory bifurcations, can be also used for examining the core region of instability due to a steady bifurcation. At the same time, it can provide directions for the implementation of passive and localized control strategies. 

The structural sensitivities of the global modes identified from global LSA are shown in Figure~\ref{fig:structural}. The maximum magnitude for the steady bifurcation lies inside the recirculation bubble. For the unsteady bifurcation computed around the axisymmetric base flow, the wavemaker region is located closer to the end of the recirculation bubble. Interestingly, for the reflectionally symmetric base flow, the sensitivity appears to move closer to the body and becomes asymmetric in accordance with the symmetry of the base flow.

\section{Direct numerical simulations}

\begin{figure}
\begin{minipage}{0.68\textwidth}
	\centering
	\includegraphics[trim={1.2cm 0cm 0cm 0cm},clip,width=1\textwidth]{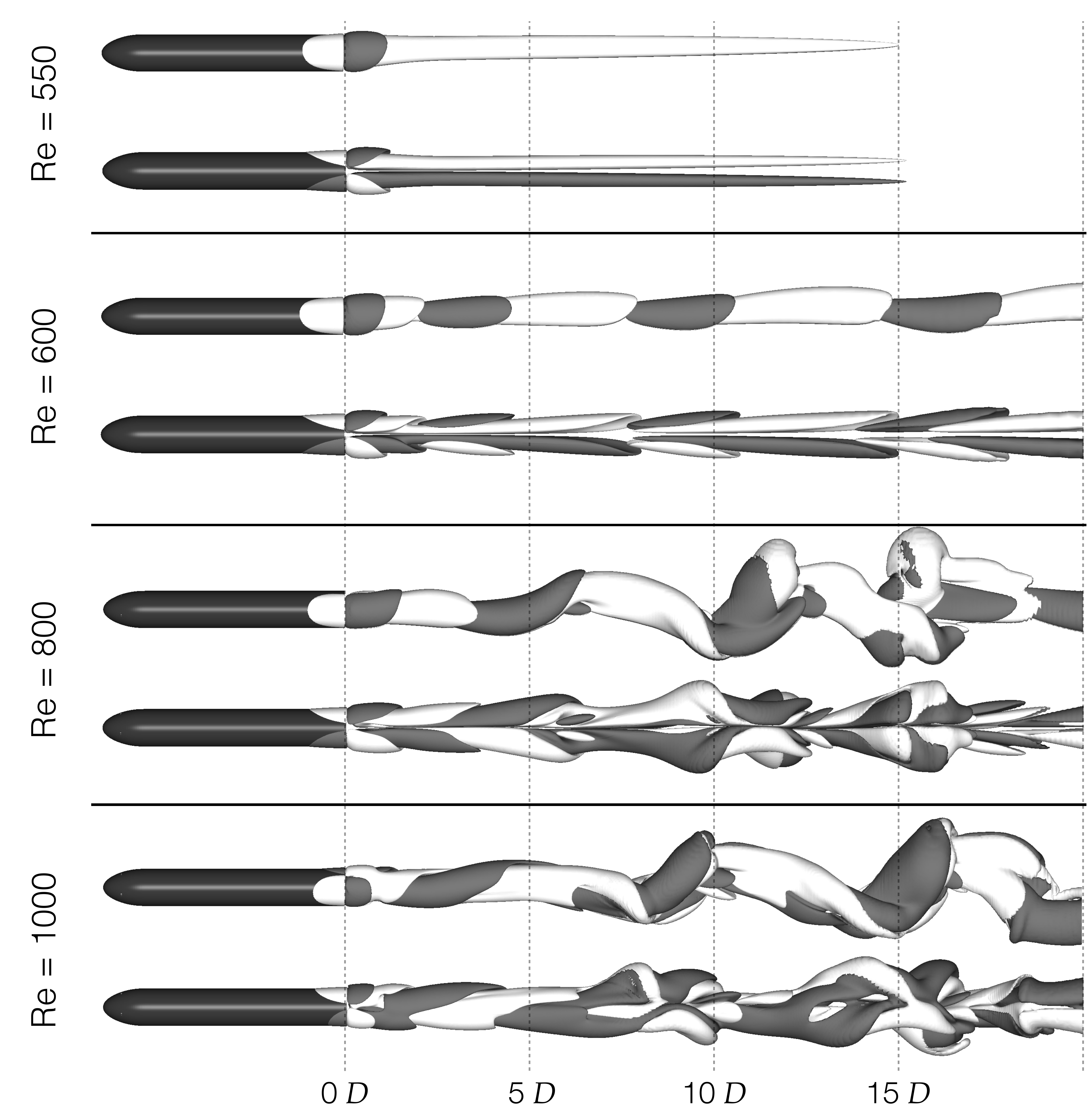}
\end{minipage}
\begin{minipage}{0.31\textwidth}
	\begin{center} 
	\vspace{1.9cm}
	\includegraphics[trim={0cm 0cm 0cm 0cm},clip,width=0.9\textwidth]{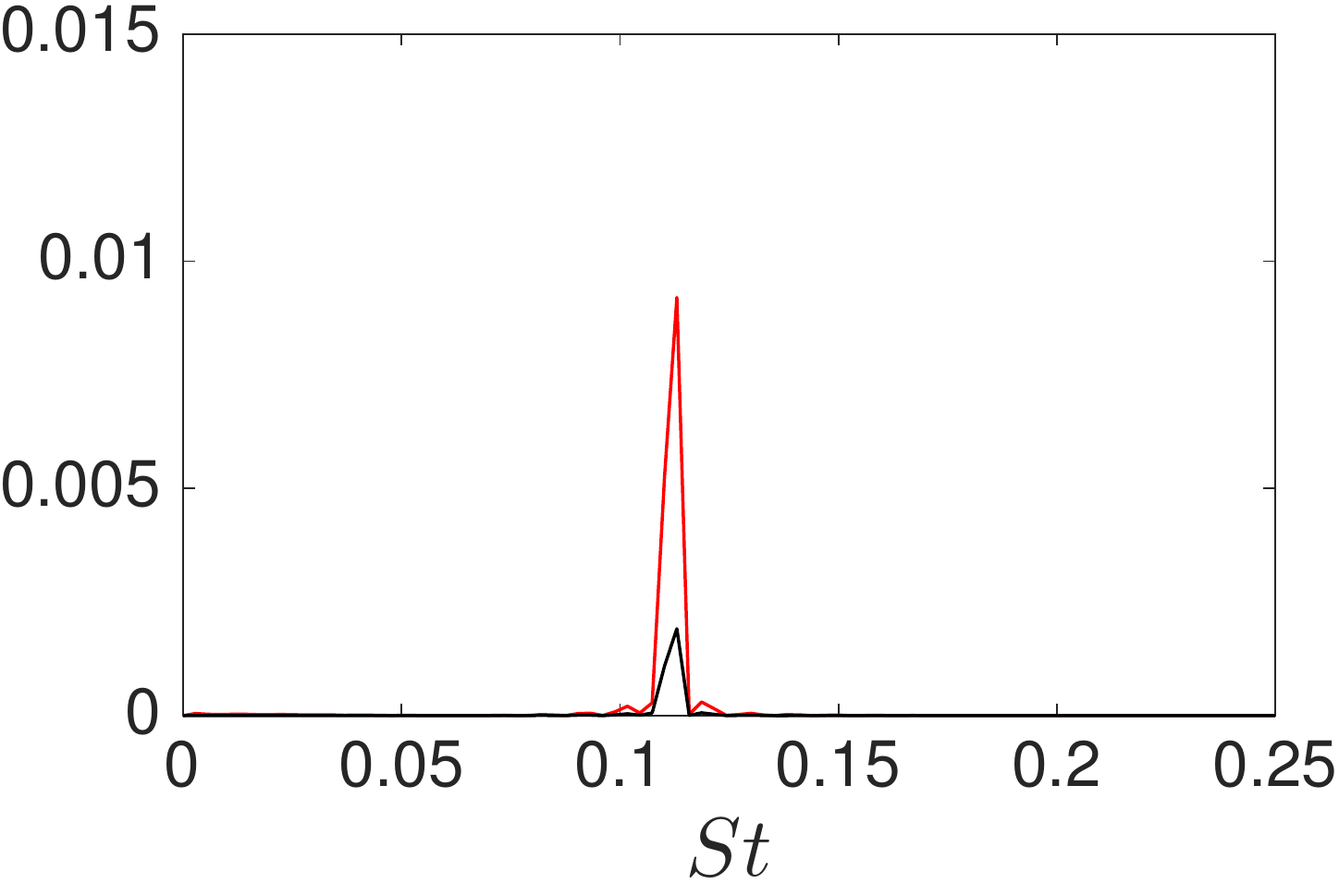} \vspace{-0.25cm} \\
	\hspace{0.0cm} \vspace{-0.2cm}
	\includegraphics[trim={0cm 0cm 0cm 0cm},clip,width=0.9\textwidth]{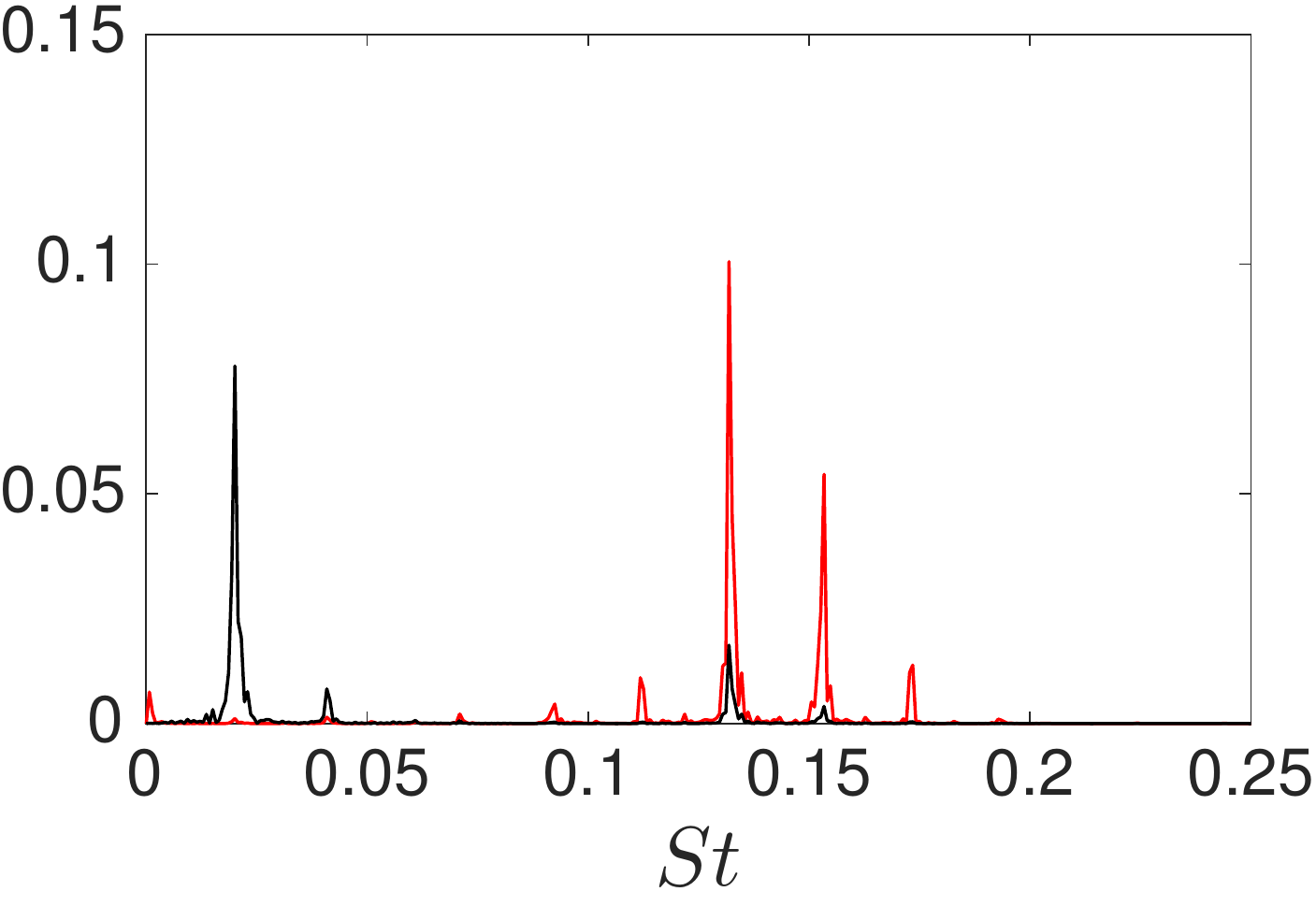} \vspace{-0.1cm} \\
	\hspace{0.0cm}
	\includegraphics[trim={0cm 0cm 0cm 0cm},clip,width=0.9\textwidth]{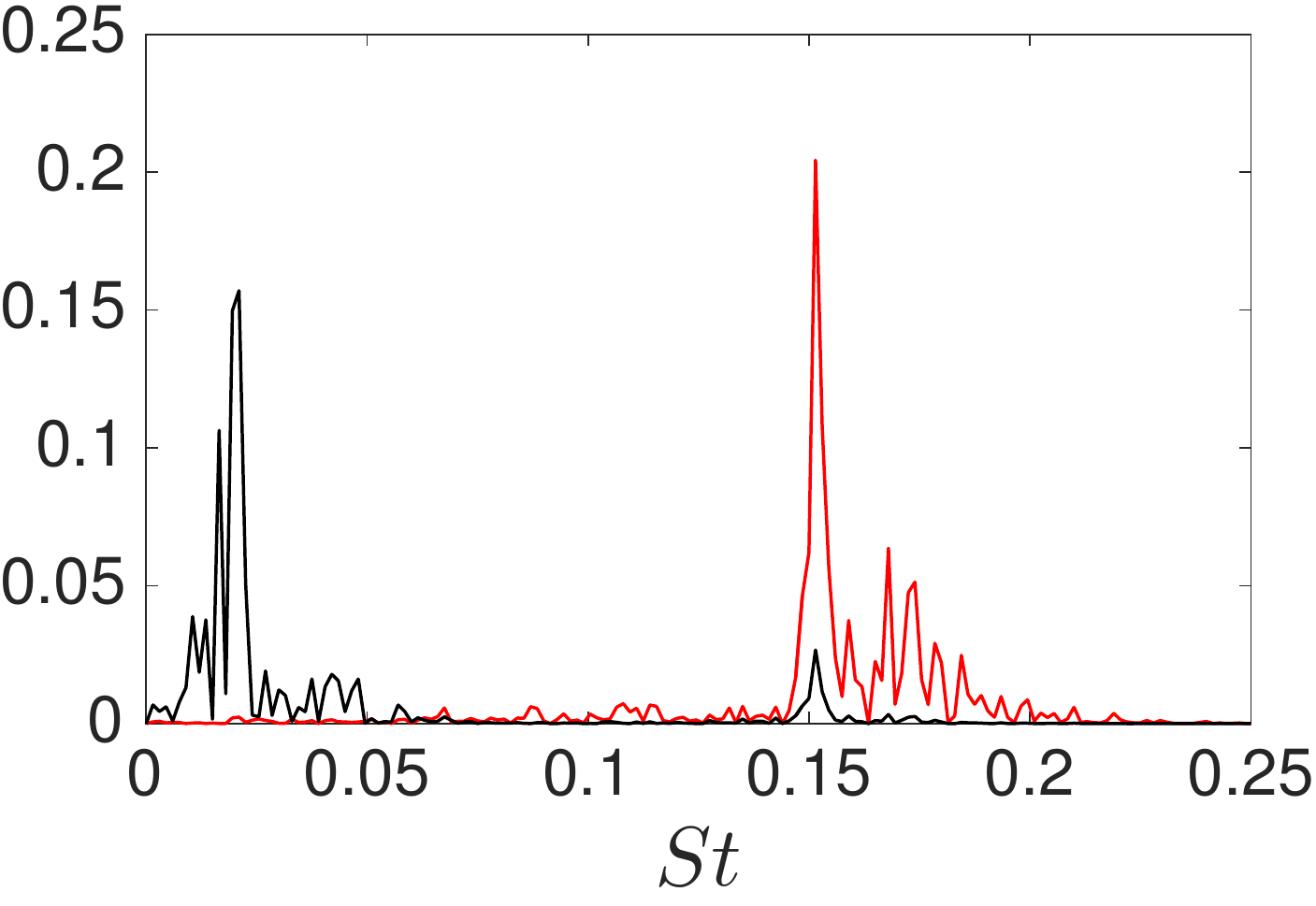} \vspace{-0.5cm} \\
	\end{center}	
\end{minipage}
	\caption{DNS simulations at $Re$: 550, 600, 800, 1000. Left: Streamwise vorticity contours, $\omega_z^*=\pm 0.05$, in the wake of the bluff-body; side (top) and plane (bottom) views. Right: Power spectral desnity (PSD) versus Strouhal number of the pressure barycenter on the base (black) and $0.5D$ downstream of the base (red) for the unsteady cases.}
	\label{fig:vorticity}
\end{figure}

DNS simulations are performed to investigate the transition stages to chaos and validate the LSA predictions. An overview of the simulations performed in terms of Reynolds number and spatio-temporal behavior is given in Table~\ref{tab:numcases}. The simulations span over the range between $Re=300$ and $Re=1500$. In Figure~\ref{fig:vorticity}, instantaneous snapshots of the normalized streamwise vorticity $\omega_z^* = \omega_z  D / w_\infty$ iso-contours for the various regimes are shown, in addition to the power spectral density (PSD) of the center of pressure at the base and half-diameter downstream. 

For $Re<400$, the base flow is axisymmetric and steady (not shown here). 
At $Re = 420$, the steady flow is asymmetric and the rotational symmetry is broken. The new flow topology is characterized by reflectional symmetry around a plane passing through the axis of the body. The angle of the symmetry plane is determined by the initial conditions. At $Re = 550$, the side view in Figure~\ref{fig:vorticity} shows that the streamwise vortices are not aligned in the streamwise direction but exhibit an increasing eccentricity as the downstream distance from the base increases. The eccentricity increases with the Reynolds number and can be used to evaluate the $Re_c$ of the steady bifurcation \citep{Bohorquez2011}.

For $Re \ge 600$, the flow is unsteady and anti-symmetric vortices are shed periodically. A dominant single frequency appears in the PSD of the center of pressure.  The LSA predictions are in agreement with the DNS.

For $Re>675$ the flow is aperiodic, presumably quasi-periodic, i.e., having two incommensurate frequencies, and multiple frequencies appear in the PSD. First, the higher harmonics of the vortex shedding become stronger. Second, a low-frequency peak at $St=0.027$ is observed. The vortex shedding becomes irregular and  bursts of vorticity occur approximately every 5 vortex shedding cycles, as shown in Figure~\ref{fig:vorticity}. Interestingly, the wake preserves the reflectional symmetry. A fully nonlinear characterization of these solutions can be obtained by advanced techniques from time-series analysis~\citep{Hegger1999}, such as phase-space reconstruction, recurrence analysis, Lyapunov analysis and entropy calculations. This is under way and will be reported in a separate paper. 

For $Re>900$, high irregularity is observed in space and time. All the spatial and temporal symmetries of the flow are broken. The PSD of the fluctuations in the near wake show rich energy content around the two main frequencies of the previous regime, and the energy of the higher vortex shedding harmonics has spread over a wide frequency range. For $Re=900$, the Lyapunov exponents for this flow are calculated in \cite{bloniganCTR2016}. One Lyapunov exponent is found positive, which indicates that the solution is low-dimensional chaotic. Furthermore, it is found that the irregular bursts observed above are associated with discrete peaks with positive finite-time Lyapunov exponent values, indicating the chaotic transition of the wake.
\begin{table}
\centering
\renewcommand{\arraystretch}{1.5} 
  {\small{\begin{tabular}{ l l | l  }
 Temporal $\;\;\;\;\;\;$& Spatial $\;\;\;\;\;\;$ & $Re$  \\ \hline
Steady 		& Axisymmetric 		& 300, 400                    \\
Steady 		& Reflectional symmetry 	& 420, 430, 500, 550 \\
Periodic 		& Reflectional symmetry 	& 600, 625, 650\\ 
Aperiodic 		& Reflectional symmetry  	& 675, 700, 750, 800 \\
Chaotic 		& No symmetry  		& 900, 1000, 1500 \\
  \end{tabular}}}
\caption{Characterization of the DNS solutions for the Reynolds numbers investigated.}\label{tab:numcases} 
\end{table}

\section{Conclusions}

We have investigated and characterized the transition to chaos of an  axisymmetric 3D bluff-body wake using global LSA and DNS. We have shown that linear stability analysis is an accurate tool to characterize the early transition and the symmetry-breaking sequences. For the axisymmetric wake,  the two initial symmetry-breaking bifurcations are associated with spatial symmetry breaking of the rotational symmetry, giving rise to a steady reflectionally symmetric wake, and temporal symmetry breaking, giving rise to unsteady vortex shedding. A fully 3D stability analysis was employed in order to capture accurately the transition from the steady reflectionally symmetric regime to single-frequency  vortex shedding.

For higher Reynolds numbers above the unsteady shedding threshold, $Re > 675$, DNS revealed that the wake preserves the reflectional symmetry. However,  intermittent bursts, which interrupt the laminar shedding,  were identified. For $Re>900$ chaotic behavior is established. The wake breaks the reflectional symmetry and random reorientations in the azimuthal direction occur.  Interestingly, the laminar symmetry-breaking instabilities persist even at high Reynolds numbers \citep{Rigas2014JFMr,Rigas2015JFMr}. Future directions of this study   involve the use of Lyapunov analysis \citep{bloniganCTR2016} to extend eigenvalue stability analysis and sensitivity to turbulent regimes.

\subsection*{Acknowledgments}
The authors acknowledge use of computational resources from the Certainty cluster awarded by the National Science Foundation to CTR. 
\bibliographystyle{ctr}
\bibliography{references}

\end{document}